\documentclass[10pt]{IEEEtran}
\ifCLASSINFOpdf
\else
\fi
\usepackage{subfig}
\usepackage{array}
\usepackage{amsmath} 
\usepackage{graphicx,wrapfig,lipsum}
\usepackage{rotating}
\usepackage{amsmath, amssymb, amsthm}
\usepackage{stmaryrd}
\usepackage{tikz}
\usepackage{verbatim}
\usepackage{url}
\usepackage[utf8]{inputenc}
\usepackage[english]{babel}

\newtheorem{corollary}{Corollary}[]

\usepackage{multicol}
\usepackage{xr-hyper}

\usepackage{algorithm}
\usepackage{algorithmicx}
\usepackage{algpseudocode}
\usepackage{algcompatible}
\usepackage{scalerel}
\usepackage{multicol}
\usepackage{setspace}
\newtheorem{definition}{Definition}
\usepackage[noadjust]{cite}
\usepackage{bbm}
\usepackage[normalem]{ulem}
\usepackage{color}
\usepackage{dsfont}
\usepackage{enumitem}
\usepackage{bm}
\usepackage[nolist,nohyperlinks]{acronym}
\renewcommand\IEEEkeywordsname{Keywords}
\usetikzlibrary{automata}
\usetikzlibrary{arrows}

\allowdisplaybreaks
\usepackage{tikz}
\usetikzlibrary{automata,positioning}
\bstctlcite{IEEEexample:BSTcontrol}
\usepackage{verbatim}

\bstctlcite{IEEEexample:BSTcontrol}
\usepackage{caption}
\usepackage{textcomp}
\usepackage{scalerel}
\usepackage{tikz}
\usetikzlibrary{svg.path}
\usepackage{float}

\usepackage{hyperref}

\definecolor{orcidlogocol}{HTML}{A6CE39}
\tikzset{
  orcidlogo/.pic={
    \fill[orcidlogocol] svg{M256,128c0,70.7-57.3,128-128,128C57.3,256,0,198.7,0,128C0,57.3,57.3,0,128,0C198.7,0,256,57.3,256,128z};
    \fill[white] svg{M86.3,186.2H70.9V79.1h15.4v48.4V186.2z}
                 svg{M108.9,79.1h41.6c39.6,0,57,28.3,57,53.6c0,27.5-21.5,53.6-56.8,53.6h-41.8V79.1z M124.3,172.4h24.5c34.9,0,42.9-26.5,42.9-39.7c0-21.5-13.7-39.7-43.7-39.7h-23.7V172.4z}
                 svg{M88.7,56.8c0,5.5-4.5,10.1-10.1,10.1c-5.6,0-10.1-4.6-10.1-10.1c0-5.6,4.5-10.1,10.1-10.1C84.2,46.7,88.7,51.3,88.7,56.8z};
  }
}

\newcommand\orcidicon[1]{\href{https://orcid.org/#1}{\mbox{\scalerel*{
\begin{tikzpicture}[yscale=-1,transform shape]
\pic{orcidlogo};
\end{tikzpicture}
}{|}}}}

	\begin{acronym}
    \acro{4G}{fourth generation}
    \acro{5G}{fifth generation}
    \acro{AoA}{angle of arrival}
   	\acro{AoD}{angle of departure}
    \acro{BCRLB}{Bayesian CRLB}
    \acro{BS}{base station}
	\acro{CDF}{cumulative density function}
    \acro{CF}{closed-form}
    \acro{CRLB}{Cramer-Rao lower bound}
    \acro{CSMA}{carrier sense multiple access}
    \acro{DSRC}{dedicated short-range communication}
    \acro{EI}{Exponential Integral}
    \acro{eMBB}{enhanced mobile broadband}
    \acro{FIM}{Fisher Information Matrix}
    \acro{GPS}{global positioning system}
    \acro{GNSS}{global navigation satellite system}
    \acro{HetNets}{heterogeneous networks}
    \acro{LOS}{line of sight}
    \acro{MBS}{macro base station}
    \acro{MIMO}{multiple input multiple output}
    \acro{mm-wave}{millimeter wave}
    \acro{mMTC}{massive machine-type communications}
    \acro{MS}{mobile station}
    \acro{MVUE}{minimum-variance unbiased estimator}
    \acro{NLOS}{non line-of-sight}
    \acro{OFDM}{orthogonal frequency division multiplexing}
    \acro{PDF}{probability density function}
    \acro{PGF}{probability generating functional}
    \acro{PLCP}{Poisson line Cox process}
    \acro{PLT}{Poisson line tessellation}
    \acro{PLP}{Poisson line process}
    \acro{PPP}{Poisson point process}
    \acro{PV}{Poisson-Voronoi}
    \acro{QoS}{quality of service}
    \acro{RAT}{radio access technique}
    \acro{RSSI}{received signal-strength indicator}
    \acro{RSU}{roadside unit}
    \acro{SBS}{small cell base station}
   	\acro{SINR}{signal to interference plus noise ratio}
    \acro{SNR}{signal to noise ratio}
	\acro{ULA}{uniform linear array}
	\acro{UE}{user equipment}
 	\acro{URLLC}{ultra-reliable low-latency communications}
    \acro{V2V}{vehicle-to-vehicle}    
    \acro{V2X}{vehicle-to-everything}
    \acro{UE}{{user equipment}}
    \acro{MEC}{{multi-access edge computing}}
    \acro{MAB}{{multi-armed bandit}}
    \acro{RL}{{reinforcement learning}}
\end{acronym}

\begin{document}

\title{An Online Algorithm for Computation Offloading in Non-Stationary Environments}

\author{Aniq Ur Rahman$^{\orcidicon{0000-0003-3685-7201}}$,~\IEEEmembership{Member,~IEEE, } Gourab Ghatak$^{\orcidicon{0000-0002-8240-4038}}$, and Antonio De Domenico$^{\orcidicon{0000-0003-1229-4045}}$
\thanks{A. U. Rahman is with the Department of Electrical Communication Engineering, Indian Institute of Science, 560012 Bangalore, India.
\newline
Email: aniqurrahman@ieee.org;}
\thanks{ G. Ghatak is with the Department of Electronics and Communication Engineering, Indraprastha Institute of Information Technology Delhi, India.\newline
Email: gourab.ghatak@iiitd.ac.in;}
\thanks{A. De Domenico is with Huawei Technologies, Paris Research Center, 20 quai du Point du Jour, Boulogne Billancourt, France.\newline 
Email: antonio.de.domenico@huawei.com.}
\thanks{Manuscript received XX XX, 2020; revised XX XX, 2020.}
}

\maketitle
\IEEEpeerreviewmaketitle

\begin{abstract}
We consider the latency minimization problem in a task-offloading scenario, where multiple servers are available to the user equipment for outsourcing computational tasks. To account for the temporally dynamic nature of the wireless links and the availability of the computing resources, we model the server selection as a multi-armed bandit (MAB) problem. In the considered MAB framework, rewards are characterized in terms of the end-to-end latency. We propose a novel online learning algorithm based on the principle of optimism in the face of uncertainty, which outperforms the state-of-the-art algorithms by up to $\sim$1~s. Our results highlight the significance of heavily discounting the past rewards in dynamic environments.
\end{abstract}


\IEEEpeerreviewmaketitle

\section{Introduction}
\IEEEPARstart{T}{he} future mobile networks will be characterized by ubiquitous coverage, ultra-low latency services, quasi-deterministic communications, and the need for extremely high data rates. In this context, a radical change consists of empowering mobile devices and \acp{BS} with data processing and storage capabilities, thereby reducing the end-to-end latency of the mobile services. This paradigm is called \ac{MEC}~\cite{Mao2017}, also known as mobile edge computing. In \ac{MEC} networks, small cells integrate computing capabilities and local cache memories to the standard \ac{RAT}. Consequently, a \ac{UE} can request a small cell to run a computational assignment on its behalf, resulting in a reduced effective latency and an increased \ac{UE} battery-life. This procedure is called \emph{task} or \emph{computation offloading}~\cite{Barbarossa2014}. Additionally, the \ac{MEC}-enabled small cells can implement proactive caching strategies to satisfy the ever growing demand for downloadable multimedia content in the mobile networks, thereby limiting the load on the transport network~\cite{Bastug2014}. The \ac{MEC} resources are often divided into three categories: communication, computing, and caching~\cite{Wang2017}.

In~\cite{elbamby2019wireless} the authors have provided a detailed overview of \ac{MEC} technology and its use-cases, particularly focusing on the services requiring low-latency and highly-reliable communications. 
Several researchers have investigated policies to determine when computation offloading is more efficient than local processing. For instance, Elbamby {\it et al.}~\cite{elbamby2017proactive} have studied the task-offloading problem formulated as a matching game, subject to latency and reliability constraints. 
More recently, computation offloading was also extended to more realistic scenarios, where system dynamics and information uncertainty is taken into consideration. For example, Liao {\it et al.}~\cite{icc_fog} have proposed a robust two-stage task offloading algorithm that integrates contract theory with computational intelligence to minimize the long-term delay of task assignment.
On the same lines, \ac{MAB} is an online \ac{RL} framework that can be used to find an optimal policy when the reward distribution of the actions is not \textit{a priori} known~\cite{lattimore2018bandit}. In particular, we focus on the case where the system characteristics, i.e., the \ac{MEC} resource availability and the wireless channel are {\textit{non-stationary}}\footnote{{This refers to a random process whose probability distribution changes in time or space \cite{besbes2014stochastic}.}}. It must be noted that, in non-stationary scenarios, off-the-shelf \ac{RL} algorithms may indeed be sub-optimal due to the usage of outdated information. Therefore, it becomes necessary to {\it forget} past rewards and rapidly update the reward distribution based on recent information. However, selecting the policy refresh rate is challenging since the \textit{agent} is typically not aware of the temporal behaviour of the system. 

Earlier, researchers have come up with the idea of \textit{discounting} the past rewards, to make the \ac{RL} system adaptive to the dynamic changes and introduced the discounted variants \cite{raj2017taming, garivier2008upper} of classical \ac{RL} algorithms.
Garivier and Moulines~\cite{garivier2008upper} considered a scenario where the distribution of the rewards remain constant over epochs and change at unknown time instants (i.e., abrupt changes). They analyzed the theoretical upper bounds of the regret for the discounted upper confidence bound (UCB) and sliding window UCB.
Gupta {\it et al.}~\cite{gupta2011thompson}, extended this idea to Bayesian methods, and proposed the {Dynamic Thompson Sampling (Dynamic TS)}.
Hartland {\it et al.}~\cite{hartland2006multi} considered dynamic bandits with {abrupt changes} in the reward generation process, and proposed an algorithm called {Adapt-EvE}. 
Slivkins and Upfal.~\cite{slivkins2008adapting} considered a dynamic bandit setting where the {reward evolves as Brownian motion} or a random walk, and provided results of regret linear in time horizon. 
Sana {\it et al.}~\cite{sana2019multi} have solved the problem of optimizing the UE-BS association by employing Deep Reinforcement Learning. 
Liao {\it et al.}~\cite{liao2019learning} have maximized the long-term throughput for a machine type device (MTD) subject to energy and data-size constraints in a learning-based channel selection framework. The learning algorithm proposed is a variant of UCB.
However, these works do not take into account, the abrupt changes at unknown times.

In this paper, we model the \ac{MEC} server selection problem as the exploration-exploitation dilemma of a restless \ac{MAB} framework with non-stationary rewards. For this problem, we propose an online learning algorithm \textit{Sisyphus} that {is model-free and is based} on the principle of optimism in the face of uncertainty. In particular, we selectively retain the knowledge of the past rewards so as to keep up with the dynamic environment. We show that Sisyphus achieves the lowest normalized regret as compared to the other algorithms proposed for the non-stationary bandit problem, namely, Thompson sampling (TS), discounted TS, discounted optimistic TS, and discounted UCB. Consequently, Sisyphus is shown to reduce the end-to-end latency by up to 1~s under the considered test environment.

\section{System Model}
{We focus on} a \ac{UE} offloading its computational task to {a nearby} \ac{MEC} server $s_i \in \mathcal{S}$, where $\mathcal{S}$ represents the set of all servers.
We assume that one task is offloaded by the UE in each time-step $t \in \{1, 2, 3, ..., T \}$ of duration $\delta$.
The aim of the \ac{UE} is to select the \ac{MEC} server which results in a minimum delay, while taking into account the task execution and signal propagation delays.

The \ac{MEC} server $s_i$ performs the task with intensity $\kappa$, which denotes the CPU cycles required to process a byte of task, using  its available computing resources, which evolves over time \cite{dandachi2019artificial}.
Unlike the centralized architecture in~\cite{liang2019multiuser}, we consider a distributed system where each user selects an MEC server independently of the other users' decision.

Specifically, the link between the \ac{UE} and the \ac{MEC} server is assumed to be affected by dynamic blockages, where the probability of blockage of the server $s_i$ is denoted by $p_{B,i}$. In addition, we model the \ac{MEC} servers as the arms in an \ac{MAB} framework, where the resource availability $a_i(t)$, varies with time in a \textit{doubly-stochastic} manner.
The computing resources available at time-step $t$ is expressed as $a_i(t)c_i$, where $c_i$ is the maximum computing capacity\footnote{Computing capacity refers to the frequency of the processor clock, i.e., number of cycles per second, typically measured in GHz.} of the server and $a_i(t) \in (0,1)$ is the fraction of the computing capacity available at time $t$. We refer to this quantity $a_i(t)$ as \textit{resource availability}.

We assume that the
 number of UEs associated with a server changes after certain number of time-steps, which in turn impacts the resource availability. This set of consecutive time-steps constitute an \textbf{epoch}.
If the probability that the number of UEs connected $v(t)$ to a server $s_i$ changes in a single time-step is $p = \text{Pr}\{v(t) \neq v(t-1)\}$, then the probability that it remains unchanged for $\Delta$ consecutive time-steps, is given by the geometric distribution~\cite{vaseghi1995state}: $$\Pi_{l=1}^{\Delta} (1-p) = (1-p)^{\Delta}.$$ We set $p = \frac{1}{\Lambda_i}$ where $\Lambda_i$ is the mean value of epoch duration. The $j^{\rm th}$ epoch size $\Delta_i^j$ can then be drawn from the distribution:
\begin{equation}
    p_{\Delta_i^j}(\Delta_i^j = \Delta; \Lambda_i) = \bigg( 1-\frac{1}{\Lambda_i} \bigg)^{\Delta},
\end{equation}
where the expected value $\mathbb{E}\{ \Delta_i^j \} = \Lambda_i$.

The instantaneous resource availability of an MEC server $a_i(t)$ is a function of the associated UEs.
If server $s_i$ can accept upto $N$ users at a time, and $q$ users offload their tasks to it, then, $a_i(t)=1-\frac{q}{N}$.  

Now, we derive the probability that $q$ UEs offload their tasks to the server at a given time-step. The considered scenario is as follows: (i) there are $w$ UEs in communication range of the MEC server, (ii) for the $j^{\rm th}$ epoch, out of these $w$ UEs, $v_j$ are connected to the small cell hosting the server, (iii) at a given time-step $t$ within the $j^{\rm th}$ epoch, out of these $v_j$ UEs, only $q_{t,j}$ UEs offload their computation tasks.\\
The probability that $v_j$ UEs out of $w$ are connected to the server follows a binomial distribution:
\begin{equation}
    p_{v_j}(v_j=v) = {w \choose v} \psi_0^v (1-\psi_0)^{w-v},
    \label{v}
\end{equation}
where $\psi_0$ is the probability of a single in-range UE to be connected to the server. The value $\psi_0$ is specific for a server $s_i$ because of the radio characteristics of the environment surrounding $s_i$ (e.g., blockages). Out of these $v_j$ UEs, only a fraction  of the UEs offload their task to the server {e.g., depending on the task computational complexity}. Therefore, we denote with $\psi_1$ the probability that a connected UE decides to offload a task. Then, $q_{t,j}$ follows a binomial distribution:
\begin{equation}
    p_{q_{t,j}}(q_{t,j}=q) = {v_j \choose q} \psi_1^q (1-\psi_1)^{v_j-q}.
    \label{q}
\end{equation}
For a given server $s_i$, the resource availability at time-step $t$ in the $j^{\rm th}$ epoch is then expressed as: $a{_i}(t) = 1 - \frac{q_{t,j}}{N}$.

Therefore, the dynamic resource availability characteristics of a server $s_i \in \mathcal{S}$ can be controlled through the parameters $\{\psi_0, \psi_1, w, N, \Lambda\}_i$.

Let us assume that the amount of uplink data {related to the task to be offloaded} be given by {$L_U$} bytes. The downlink data {size, after the MEC server processing, is denoted as $L_D$ and is related to the uplink data as: $L_D = \Omega  L_U, \Omega \in \mathbb{R}^+$}. Furthermore, let $\gamma$ denote the path-loss exponent of the transmissions, which varies depending on the blockage conditions, i.e., whether the channel visibility state is in \ac{LOS} or \ac{NLOS}. Additionally, let the reference uplink \ac{SINR} at 1 m be denoted as $\mathcal{P}_U$. {Similarly, the downlink \ac{SINR} at 1 m is denoted as $\mathcal{P}_D$. The uplink and downlink bandwidths are denoted as $B_U$ and $B_D$ respectively}. Thus, the total transmission delay $\tau_i(t)$ {when the distance between the UE and server is $r_i$,} can be written as:

\begin{equation}
    \tau_i(t) = \sum_{Z \in \{U, D\}} \frac{L_Z}{B_Z \log_2(1 + \mathcal{P}_Z\cdot r_i^{-\gamma})};
\end{equation}

For the processing phase, the computation delay $\eta_i(t)$ is defined as the time taken by the \ac{MEC} server $s_i$ to process the data and generate the output, { which is expressed mathematically as:} $$\eta_i(t) = \frac{\kappa L}{c_i a_i(t)}.$$

Then, the total delay is the sum of transmission and computation delays: $D_i(t) = \tau_i(t) + \eta_i(t)$.
Finally. the reward associated with server $s_i$ at time-step $t$ is denoted by $\rho_i(t)$. {Let $D_{\max}$ be the latency requirement of the task that the UE wants to offload; then, we can define} the reward $\rho_i(t)$ as:
$$\rho_i(t) = \mathds{1}_{ \{ D_i \leq  D_{\max} \} },$$ 
which ensures that the reward is positive and bounded by 1. The UE follows a policy $\pi$ (see Section \ref{sec: Algo}) to select an arm at each time-step. Let $\rho_j({t})$ be the reward of the arm chosen at time-step $t$ and $\max \rho_i({t})$ denote the highest reward among all arms' reward; then, the \textit{time-normalized cumulative regret} $R_{\alpha}({T})$ for ${T}$ time-steps is defined as the cumulative sum of the difference between the rewards of the best arm and the chosen arm (according to $\pi$) divided by the count of time-steps ${T}$. We refer to it as the normalized regret, given by:
\begin{equation}
    R_{\pi}({T}) =  \frac{1}{{T}}\sum_{{t=1}}^{{T}} \left(\max\limits_{s_i \in \mathcal{S}} (\rho_i({t})) - \rho_j({t}) \right).
\end{equation}
The objective of the \ac{MAB} framework is to design the policy $\pi$ so as to minimize $R_{\pi}({T})$. In the next section, we propose one such policy which outperforms the state-of-the-art \ac{MAB} algorithms.

\section{Proposed Online Learning Algorithm}
\label{sec: Algo}
We consider an $|\mathcal{S}|$-armed bandit, where the UE, at each time-step, plays the arm (i.e., selects the server) which has the highest expected reward, based on the past experiences of playing the arms. {Specifically, for each server $s_i \in \mathcal{S}$}, the UE tracks the total number of times each arm has been played, denoted by $k_i$ {and} maintains a score $\mu_i$, as described below.
\begin{definition}
{On playing the arm $s_i$ for the $k_i^{\rm th}$ time, we obtain a \textbf{reward} $\rho_i(k_i)$, then the \textbf{score} $\mu_i(k_i)$ for that arm is updated as:}
\begin{equation}
    \mu_i(k_i) = \frac{1- \alpha^{k_i -1}}{2 -\alpha  - \alpha^{k_i}} \mu_i(k_i -1) + \frac{1- \alpha}{2 -\alpha - \alpha^{k_i}}\rho_i(k_i),
    \label{learneq}
\end{equation}
where $\alpha\in [0,1)$ is the \textbf{retention rate}.
\end{definition}

The parameter $\alpha$ controls the amount of memory in the \ac{MAB} framework. Two extreme states can be determined {in the system}, by substituting the value of $\alpha=0$ and $\alpha \rightarrow 1$. If $\alpha$ is set to zero, the UE gives equal weight-age to the new reward compared to the weighted sum of previous rewards. For $\alpha \rightarrow 1$, past rewards have {a larger} effect on the current score and thereby influence more the UE's decision. In essence, lower the value of $\alpha$, the lesser memory the system has about the past rewards.

\begin{corollary}
The score assigned to arm $s_i$ can be expressed as a weighted sum of rewards, where $\phi_{\alpha}(k_i, m)$ denotes the \textbf{memory weight} for the reward when the arm $s_i$ is played for the $m^{\rm th}$ time:
    \begin{equation}
        \mu_i(k_i)= \sum_{m=1}^{k_i} \phi_{\alpha}(k_i, m)\cdot \rho_i(m); \quad k_i > 0.
        \label{learn_weight}
    \end{equation}
    {
    \begin{equation}
        \phi_{\alpha}(k_i, m) = \frac{1- \alpha}{2 - \alpha -\alpha^m} \cdot
        \left( \prod_{j = m+1}^{k_i} \frac{1- \alpha^{j-1}}{2 - \alpha -\alpha^j} \right).
        \label{learn_weight}
    \end{equation}
    }
\end{corollary}

\begin{figure}[h!]
    \centering
    \includegraphics[width=0.8\columnwidth]{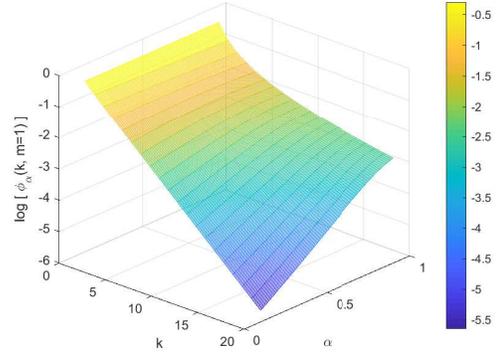}
    \caption{Memory weight across different retention rates.}
    \label{fig:phi_3d}
\end{figure}

After a certain {play-count} $k_i$, the reward at the $m^{th}$ {play}, becomes negligible {($m < k_i$)}. This is bound to happen, as the recorded reward successively fades, until it no longer affects the score of that arm. This is more intuitive than resetting the previous rewards to zero at regular predefined intervals (e.g., see~\cite{aniq_wmlc}), since smooth transitions allow to take care of abrupt changes in the reward distribution. Refreshing the score to zero at fixed intervals may either reset it too early, or too late, resulting in sub-optimal performance.
In essence, {the UE gives} importance to the score of an arm and the number of times it has been played. This prevents us from getting biased by the performance of an arm in a few trials. This is the optimistic approach, where we expect that a poorly performing arm might perform well in the future draws owing to the uncertain behaviour of the arms.
We have depicted the concept of memory weights graphically in Fig.~\ref{fig:phi_3d}. For smaller values of retention rate $\alpha$, the reward recorded for an arm fades quickly as it is played more number of times ($k$). On the other hand, for values of $\alpha$ close to 1, the reward fades slowly in comparison.

The {proposed} algorithm \textbf{Sisyphus} (SSPH) is described in Algorithm~1. The scores $\mu$ and counts $k$ for all the arms are initialized to zero in step 1 and step 2 respectively. A time loop starts in step 3 which is terminated in step 9 within which, the following operations are performed sequentially: an expected reward $\theta$ is drawn from the normal distribution (step 4) and the arm with the maximum expected reward is chosen to be played (step 5). The {play-}count of that arm (which tracks the number of times the arm has been played) is incremented by 1 (step 6). When the selected arm is played, the actual reward is revealed, after which we update the score of the chosen arm in step 7 and that of {the set of} the never-played arms {$\mathcal{S}^0$} in step 8.

\begin{figure*}
\centering
\begin{minipage}[b]{.3\textwidth}
    \centering
    \includegraphics[width=\columnwidth]{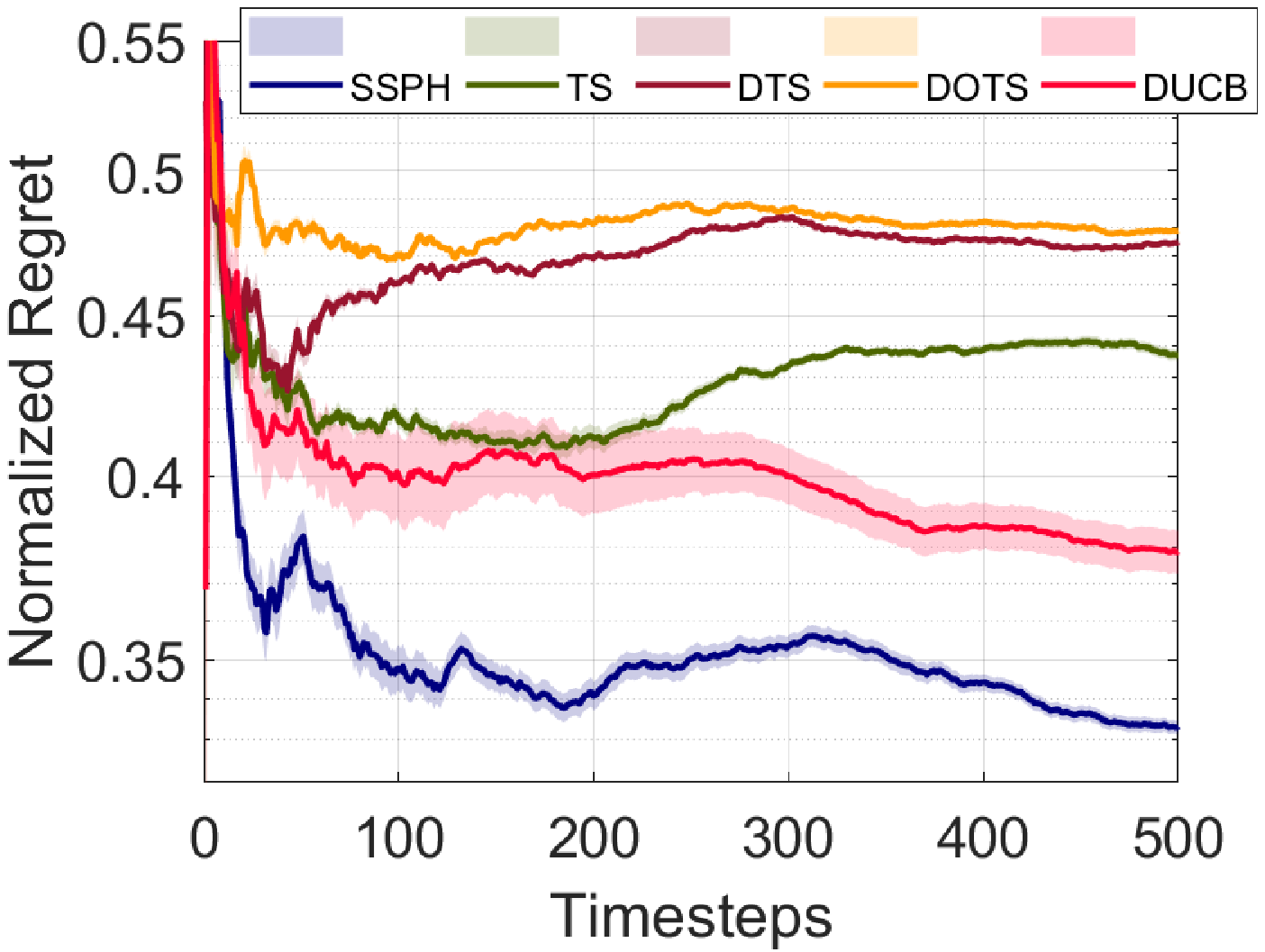}
    \caption{Normalized Regret against time. $|\mathcal{S}| = 5$.}
    \label{fig:regret}
\end{minipage}\qquad
\begin{minipage}[b]{.3\textwidth}
    \centering
    \includegraphics[width=\columnwidth]{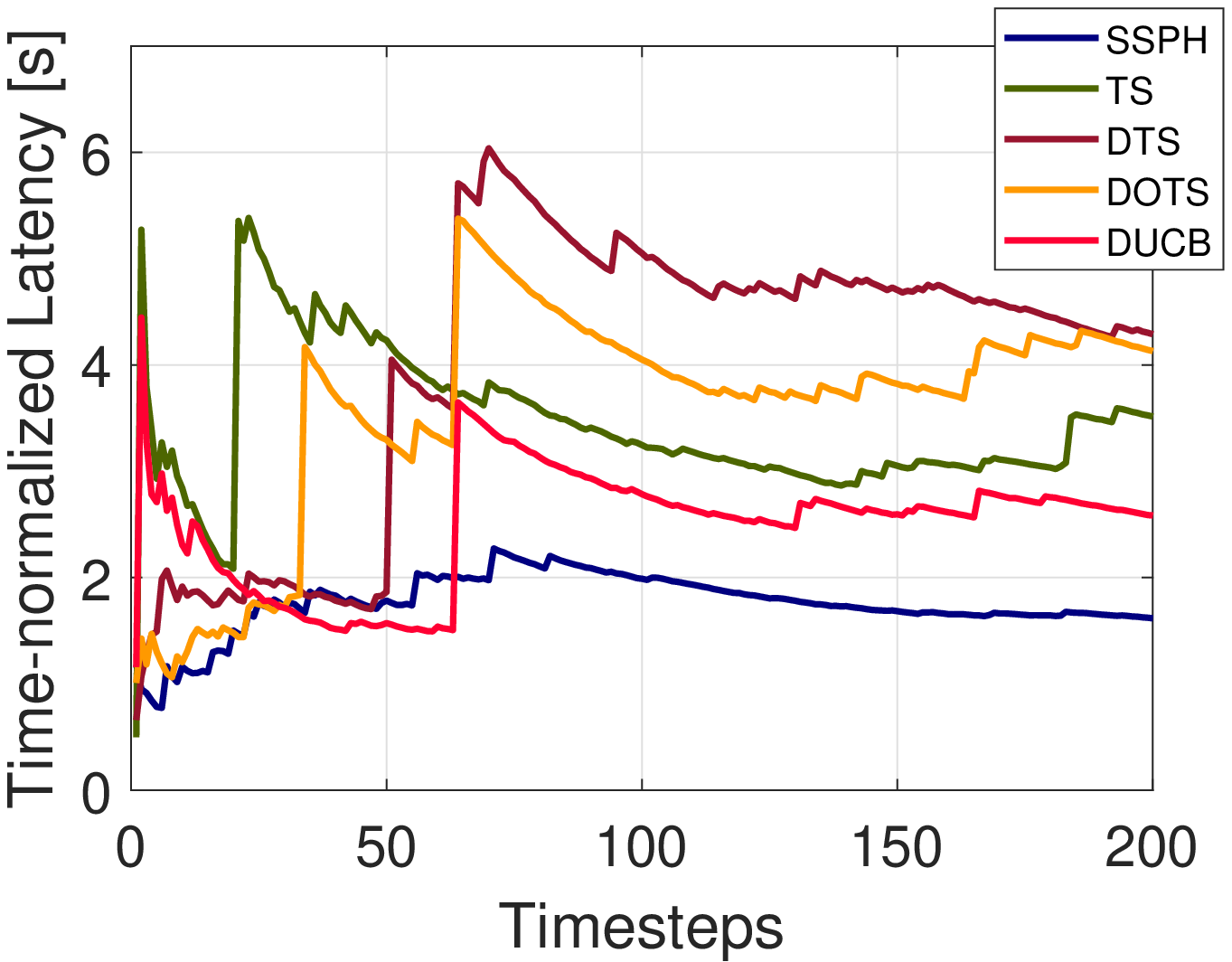}
    \caption{Normalized Latency evolution. $|S|= 5$.}
    \label{fig:latency}
\end{minipage}\qquad
\begin{minipage}[b]{.3\textwidth}
    \centering
    \includegraphics[width=\columnwidth]{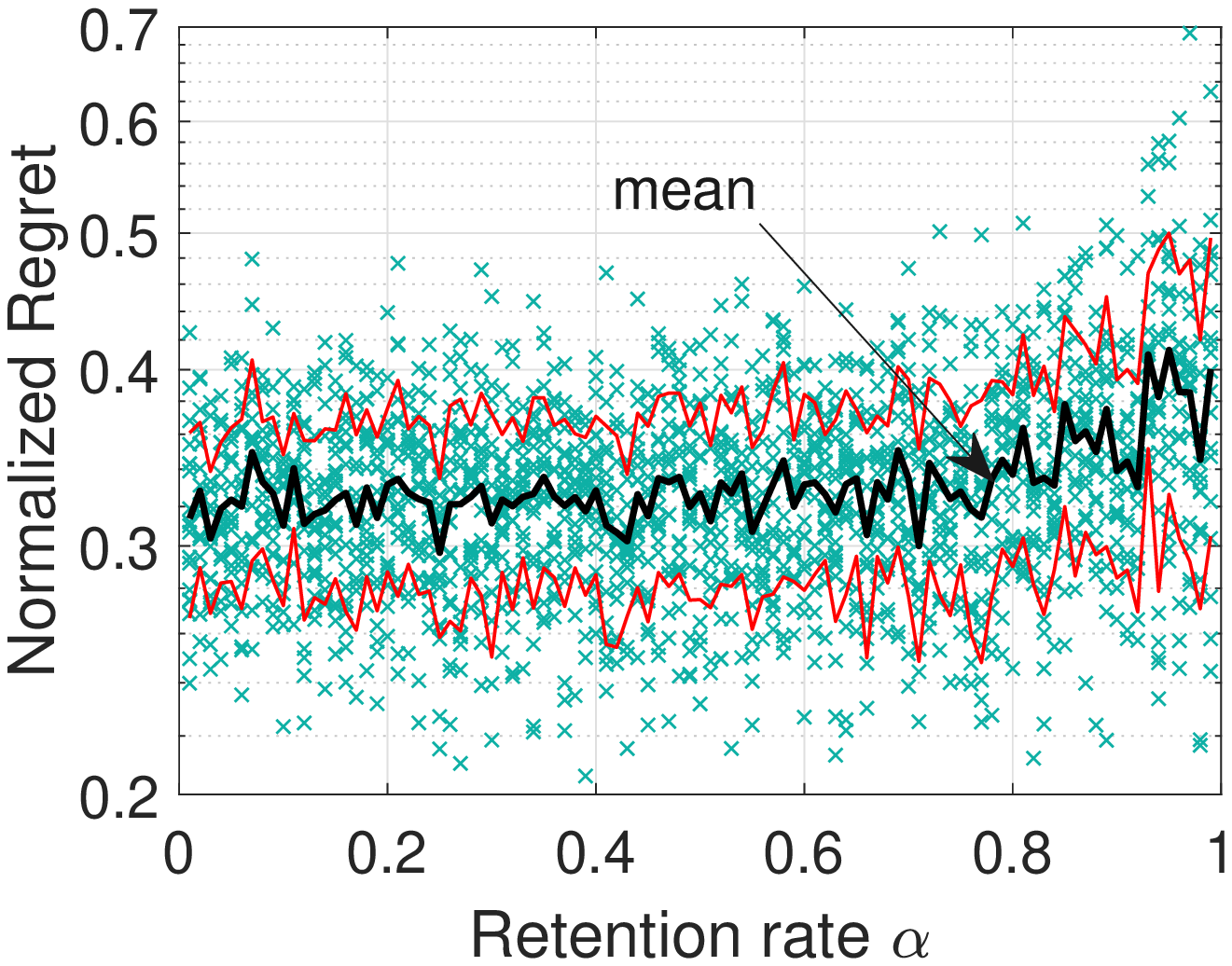}
    \caption{Normalized Regret vs. Retention rate. $|S|= 5$.}
    \label{fig:alpha_scatter}
\end{minipage}
\vspace{-10pt}
\end{figure*}

\begin{algorithm}[h!]
{Retention rate $\alpha$}
    \begin{algorithmic}[1]
    \STATE $\mu_i(0) \gets 0; \quad \forall s_i \in \mathcal{S}$
    \STATE $k_i \gets 0; \quad \forall s_i \in \mathcal{S}$
    \FOR{$t \in 1, ..., T$}
    \STATE $\theta_i \sim \mathcal{N}(\mu_i(k_i), \sigma^2); \quad \forall s_i \in \mathcal{S}$
    \STATE $s_j(t) \gets \arg \max_{s_i \in \mathcal{S}} \theta_i$
    \STATE $k_j \gets k_j +1$
    \STATE $\mu_j(k_j) \gets \frac{1- \alpha^{k_j -1}}{2 -\alpha  - \alpha^{k_j}} \mu_j(k_j -1) + \frac{1- \alpha}{2 -\alpha - \alpha^{k_j}}\rho_j(k_j)$
    \STATE $\mu_u(0) \gets \frac{1}{|\mathcal{S} \setminus \mathcal{S}^0|}\sum_{s_i \in \mathcal{S} \setminus \mathcal{S}^0} \mu_i(k_i); \quad \forall s_u \in \mathcal{S}^0 $
    \ENDFOR
    \end{algorithmic}
    \caption{SISYPHUS}
    \label{sisyphus_algo}
\end{algorithm}

The algorithm is based on the principle of optimism in the face of uncertainty\footnote{The optimism in the face of uncertainty principle states
that the actions should be chosen assuming the environment to be as nice as plausibly possible.}~\cite{lattimore2018bandit}. We first assign the score of zero to each arm and then draw the {expected} reward from a normal distribution with mean equal to the score $\mu_i(k_i)$ and variance\footnote{The appropriate value of $\sigma^2$ can be tuned based on empirical history.} equal to $\sigma^2$. 
This is a Bayesian approach~\cite{poupart2006analytic} and allows us to look for expected rewards in the neighborhood of the recorded score $\mu_i(k_i)$, since it is not wise to make decisions by comparing the scores of the arms directly, in a non-stationary environment.
This enables us to predict values which would otherwise be ignored in a greedy technique~\cite{wunder2010classes}.
As we play, we update the score of the arms that have never been sampled as the average of the scores of the played arms. This boosts the probability of exploration of the unexploited arms. In contrast to the classical MAB algorithms, e.g., UCB, which add specific terms to facilitate exploration, the proposed scheme is a randomized algorithm in which the exploration-exploitation trade-off is based on a Bayesian framework.

In the following section, we show several numerical results that compares our algorithm with other state-of-the-art algorithms.

\section{Simulation Results}
To assess the proposed online learning algorithm, we define five classes of servers $\{s_1, s_2, s_3, s_4, s_5 \} \in \mathcal{S}$, whose characteristics are described in Table~\ref{tab:servers}. In our simulations, the $j^{\rm th}$ server is assigned to one of these classes of servers as: $s_j \gets s_{j \pmod 5}; \quad j>5$, where $\mod$ denotes the modulus operation.

\begin{table}[h]
    \centering
    \begin{tabular}{| >{\small}p{1.5cm}| >{\small}p{0.8cm} | >{\small}p{0.8cm}| >{\small}p{0.8cm} | >{\small}p{0.8cm}| >{\small}p{0.8cm} |}
    \hline
    \textbf{Parameter} & $s_1$ & $s_2$ & $s_3$ & $s_4$ & $s_5$\\
    \hline
    ${\psi_0}$ & 0.7 & 0.6 & 0.5 & 0.4 & 0.3 \\
    $\Lambda_i$ & 100 & 150 & 100 & 100 & 50 \\
    $r_i$ [m] & 7 & 10 & 12 & 14 & 16 \\
    $p_{B,i}$ & 0.3 & 0.4 & 0.5 & 0.6 & 0.7 \\
    $c_i$ [GHz] & 5 & 3.3 & 3.3 & 3.3 & 5\\
    \hline
    \end{tabular}
    \caption{MEC Servers}
    \label{tab:servers}
\end{table}

Additional simulation parameters are:  {$L_U = 20$~MB~\cite{qi2016quantifying}, $\Omega=1$, $B_D = B_U = 500$~MHz, $\mathcal{P}_U= 20$~dBm, $\mathcal{P}_D= 40$~dBm,} $D_{\max} = {1}$~s, $\kappa = 10$ cycles/byte, $\gamma_{\text{LOS}} = 2, \gamma_{\text{NLOS}} = 4$  {$w=N=100$, $\psi_1 = 0.5 \,\forall s_i \in \mathcal{S}$, {and $\delta = 1$~s}}.

We compare the performance of Sisyphus (SSPH) with the following algorithms: Thompson Sampling (TS)~\cite{thompson1933likelihood}, Discounted Thompson Sampling (dTS)~\cite{raj2017taming}, Discounted Optimistic Thompson Sampling (dOTS)~\cite{raj2017taming} and Discounted UCB (D-UCB)~\cite{garivier2008upper}. It is important to note that the last three algorithms (dTS, dOTS, and D-UCB) are designed to tackle the issue of dynamically changing environments in the \ac{MAB} framework. The value of $\alpha$ is set to $0.6$ for Sisyphus. The discounting factor of the benchmark algorithms are chosen for their best performance: dTS ($0.8$), dOTS ($0.7$) and D-UCB ($0.5$).

\subsection{Normalized Regret}
In Fig.~\ref{fig:regret}, we plot the temporal evolution of the normalized regret for the different algorithms. Here, the solid lines represent the mean of the normalized regret, and the shaded region represents the variance. The proposed algorithm SSPH evidently outperforms all the other algorithms and has a much lower normalized regret {$(\sim 0.32)$} compared to the other algorithms {$(> 0.37)$}. Interestingly, we observe that SSPH also has a considerably lower variance, which indicates that it is more robust than the other contending algorithms.

\subsection{Latency}
Naturally, the reduced normalized regret will be reflected on the latency performance with different algorithms. To validate this, we plot the variation of time-normalized latency for various algorithms in Fig.~\ref{fig:latency}. We observe that as the temporal process evolves, the latency of most of the contending algorithms increases gradually and settle into a higher value {$> 2.5$~s}. On the contrary, the latency of the proposed algorithm is considerably lower {($\sim 1.5$~s)}.

\subsection{Parameter Tuning}
Indeed the performance of SSPH will depend on the agility of the environment change, and the corresponding choice of $\alpha$. 
However, the algorithm developed is model-free, and takes the rewards as input at each time-step to update its score for the respective arm. The performance can be tweaked by tuning the parameters $\alpha$ which denotes how strongly the algorithm retains the past rewards and the variance $\sigma^2$ which controls the degree of exploration. For a highly dynamic system, the past rewards need to be forgotten quickly and in an environment with less number of arms, the exploration factor can be kept low.
In our work, for all the algorithms, the corresponding retention parameters are tweaked to obtain the best performance. In Fig.~\ref{fig:alpha_scatter}, we show how the normalized regret varies with varying $\alpha$ for SSPH with 5 classes of servers. Here, the {blue} scattered points are observations, black solid line is mean of the observations, {red} lines are the standard deviation around the mean value. It can be observed that the mean of the scattered point remains reasonably flat, i.e., ranging within {$[0.3, 0.35]$ for $\alpha \in [0.1, 0.6]$}. This indicates that a fairly robust selection of $\alpha$ can be made for deploying SSPH in the UE.

\subsection{Scalability}
\begin{figure}[h]
    \centering
    \includegraphics[width=0.8\columnwidth]{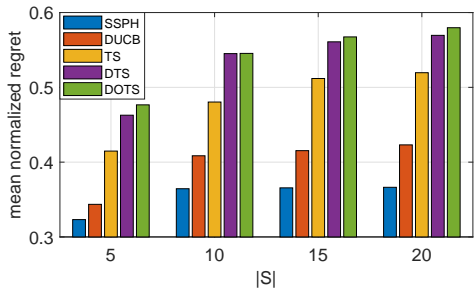}
    \caption{Mean Normalized Regret across various algorithms for different number of arms.}
    \label{fig:algo_comp}
    \vspace{-0.1cm}
\end{figure}

Next, in Fig.~\ref{fig:algo_comp}, we vary the number of arms (i.e., the number of MEC servers) $|\mathcal{S}|$ and compare the mean normalized regret of the different algorithms. {The normalized regret for TS, DTS and DOTS increases with increase in $|\mathcal{S}|$. On the other hand the normalized regret of SSPH and D-UCB does not change significantly with increase in $|\mathcal{S}|$. It must be noted that} SSPH maintains the minimum value of mean normalized regret among the contenders.

To capture these nuances of SSPH in a more concrete manner, currently we are investigating the theoretical regret bounds of the proposed algorithm and testing it for other online learning use cases.

\section{Conclusion}
In this paper, we proposed an online learning algorithm for the MAB framework with an objective to minimize the end-to-end latency in offloading computation tasks to MEC servers. In particular, we showed that selective retention of past rewards is necessary to tackle temporally varying environments. The proposed algorithm (Sisyphus) works on the principle of optimism in the face of uncertainty, and outperforms the other state-of-the-art algorithms for non-stationary MAB frameworks. We show that the proposed algorithm, in the test environment achieves a latency which is  {at least $\sim 1$~s} lower than the other benchmark algorithms.

\bibliography{bare_jrnl}
\bibliographystyle{IEEEtran}

\end{document}